\documentclass[twocolumn,showpacs,10pt,prb,floatfix,amssymb]{revtex4-1}
\usepackage{graphicx}

\begin{document}

\newcommand{\nx}{\textrm}

\title{Mixing of Edge States at a Bipolar Graphene Junction}
	
\author{H. Schmidt}
\author{J.~C. Rode}
\author{C. Belke}
\author{D. Smirnov}
\author{R. J. Haug}
\affiliation{Institut f\"ur Festk\"orperphysik, Leibniz
Universit\"at Hannover, Appelstr. 2, 30167 Hannover, Germany}

\date{\today}
\begin{abstract}

An Atomic Force Microscope is used to locally manipulate a single layer graphene sheet. Transport measurements in this region as well as in the unmanipulated part reveal different charge carrier densities while mobilities stay in the order of $10^4$~cm$^2$(Vs)$^{-1}$. With a global backgate, the system is tuned from a unipolar n-n' or p-p' junction with different densities to a bipolar p-n junction.
Magnetotransport across this junction verifies its nature, showing the expected quantized resistance values as well as the switching with the polarity of the magnetic field. The mixing of edge states at the p-n junction is shown to be supressed at high magnetic fields. \end{abstract}


\maketitle
Graphene exhibits outstanding electronic properties \cite{Novoselov2004} including high mobilities even at room temperature and a bandstructure with the valence and conduction band touching at the Dirac points. This zero bandgap makes it possible not only to continuously tune the charge carrier density, but also to change the type of majority charge carriers from electrons to holes. To achieve regions with different densities including p-n junctions, a variety of techniques can be used including topgates \cite{C_Marcus_Science,pnarray} and chemical doping \cite{chem,chem2} of defined regions, showing interesting new physics like electron-hole interference\cite{dimi} and snake states along such junctions \cite{snakestates}. One way to gain a better insight on the properties of these systems are multiterminal magnetotransport experiments and the study of the equilibrium of the edge states at the junction\cite{equilib}.\\
The Atomic Force Microscope (AFM) has proven itself as a reliable tool to create low-dimensional systems like quantum dots out of epitaxial grown 2D electron systems by scratching or local oxidation. Recent works have applied these techniques to graphene \cite{oxi1,oxi2,oxi3,bart}. It can also be used to clean and flatten the surface of graphene and thereby reach higher mobilities \cite{cleaning1, cleaning2}. Here we use the AFM to alter the electronic structure of graphene in a defined region, leading to a local doping of our device while conserving the transport properties of the graphene. The so created junction of areas with different charge carrier densities is examined using magnetotransport, showing the quantized resistance values as expected from theory.\\
\begin{figure}[t]
\includegraphics[width=.95\columnwidth]{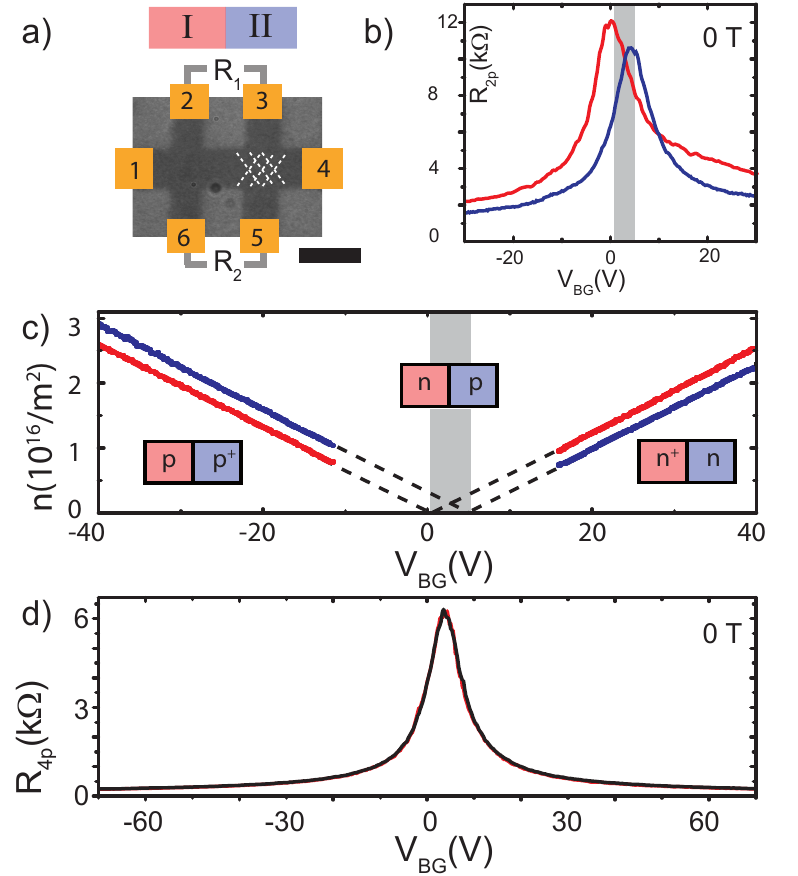}
\caption{\label{fig:bild1}a) Optical picture of the graphene Hall bar with the used contacts, the black scale bar corresponds to 5~$\mu$m. The white dotted lines indicate the manipulated area. b) Two terminal resistances in the two regions (region I red, region II blue). c) Charge carrier densities in the two parts obtained by four terminal Hall measurements. The insets show the polarities in the different regimes. d) Resistance across the junction measured at the two sides of the sample (red R$_1$, black R$_2$) at zero magnetic field.}
\end{figure}\\
The graphene sample was prepared by the standard exfoliation technique \cite{Novoselov2004} of natural graphite, and placed on top of an n-doped silicon wafer with a 330~nm thick layer of silicondioxide. Optical microscopy and analysis of the optical contrast is used to identify monolayer graphene \cite{blake}. The so selected sample is divided into two areas, I and II, as shown in Fig. 1a). While region I remains unchanged, region II is manipulated with an AFM using a diamond coated tip (Nanosensors\texttrademark DT-NCHR) in contact mode. It is moved multiple times over the surface with an applied force of F$= 8~\mu$N and a velocity of $v_{tip}=10^{-6}$ms$^{-1}$. 
As shown in the transport measurements, this procedure induces local doping to the altered part, but does not change mobilities significantly. We assume that cuts are induced due to the strong force on the hard tip. A self healing process as described by Zan et al.\cite{healing} removes the cuts but leaves behind local lattice defects. The effect of selective cleaning\cite{cleaning1} can be ruled out, since before the application of the high forces to only one part all areas have been scanned multiple times with a lower force. After the nanomachining, standard ebeam lithography is used to etch a Hall bar and apply chromium gold contacts.\\

Figure 1a) shows an optical picture of the structured graphene and the used contacts, drawn in yellow. Before the electrical properties of the sample are investigated at low temperatures, it is annealed to remove residues of the preparation process.
\begin{figure}[th]
\includegraphics[width=.95\columnwidth]{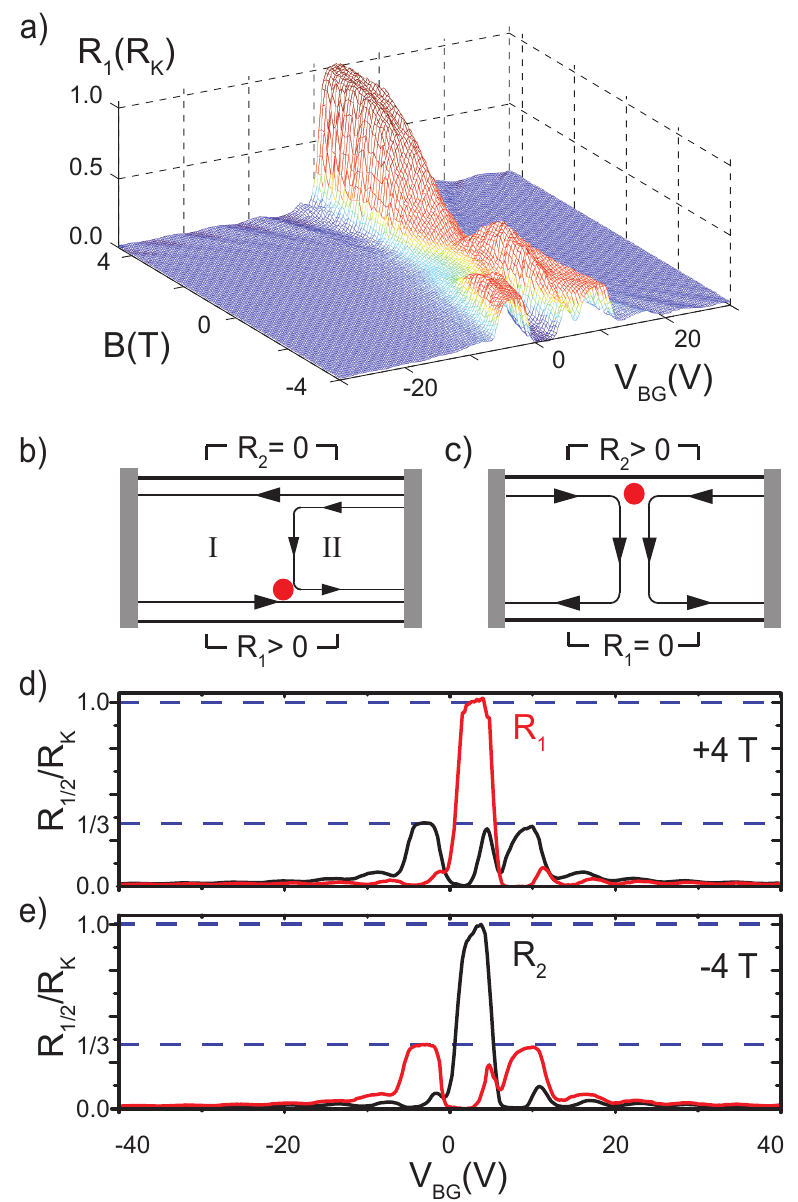}
\caption{\label{fig:bild2} a) Four terminal measurement of the longitudinal resistance R$_1$ over the junction as a function of backgate voltage and magnetic field. b) Sketch of the edge channels in the sample for the case of same polarity in both regions. The red dot marks the area of maximum non-equilibrium between the channels. c) The case for opposite polarity, i.e. a p-n junction. d) e) Resistance measurements at the two opposite sites for magnetic fields of different polarity. The blue dotted lines indicate values of 1 and 1/3 in units of the von Klitzing constant.}
\end{figure}
To characterize the two areas independently, two-terminal measurements between the contacts 2-6 and 3-5 for area I and II, respectively, are performed. The obtained field effect measurements (Fig. 1b) of the resistance show a shift of the charge neutrality point (CNP). While for the unmanipulated region I it is found at 0.2~V as expected for clean graphene, the one for the altered region II shows a shift to 5.5~V. Taking into account the thickness of the silicon dioxide, this corresponds to a difference in charge carrier density of $\Delta n= 3.5\cdot10^{15}$m$^{-2}$. The shape of both curves is similar indicating comparable mobilities in the altered and unaltered part. Figure 1c) shows the charge carrier densities in the two parts, obtained from Hall measurements performed at $B=$2~T with the current driven from contact 1 to 6 and the voltage for the two regions measured perpendicular. Both curves exhibit the same slope, but the extrapolated zero value, i.e. the CNP, is shifted. This shows, that a constant difference $\Delta n$ is present over the whole range of charge carrier densities used in these experiments. Due to this difference the system can be tuned into different states with an applied backgate voltage $V_{BG}$. For large absolute voltages, both regions exhibit the same kind of majority charge carriers, but with different densities. For 0.2~V$\leq V_{BG}\leq$ 5.5~V, a p-n junction is formed with electrons in region I and holes in region II.\\
In the absence of a magnetic field (Fig. 1d) the two longitudinal resistances $R_1$ (contacts 2-3) and $R_2$ (contacts 5-6), measured in four terminal setup across the junction, exhibit similar values with an averaged field-effect mobility of $\mu\approx$~14000 cm$^2$(Vs)$^{-1}$. These values are comparable to the ones in unmanipulated single layer graphene, indicating that the manipulation did not alter the transport scattering rates significantly.\\
To characterize the junction further, magnetotransport measurements are performed at $T=1.5$~K. Figure 2a) shows the longitudinal resistance R$_1$ as a function of backgate voltage and magnetic field. Interestingly, there is a strong dependence on the polarity of the magnetic field, which is further discussed in the following.\\The transport in the quantum Hall regime can be well understood by the edge-channel picture \cite{haugreview}. The different situations are sketched in Fig. 2. For the case of same polarity (Fig. 2b) channels being present in both regions travel across the sample while the additional ones due to a higher carrier density and therefore filling factor $\nu=nh/eB$ circulate in only one region. This leads to different longitudinal resistances which are described by the Landauer-B\"uttiker formalism as fractions of the von Klitzing constant R$_K$\cite{pntheo,pn1,equilib}:
\begin{equation} R_1=0\quad\quad R_2=R_K(\frac{1}{|\nu_1|}-\frac{1}{|\nu_2|})\quad.\end{equation}
The other case with different polarities, i.e. a p-n junction, the equilibration of the counterpropagating edge states leads to longitudinal resistances of
\begin{equation} R_1=R_K(\frac{1}{|\nu_1|}+\frac{1}{|\nu_2|}) \quad\quad R_2=0\quad,\end{equation}
if the coupling between the two regions is perfect, i.e. very strong mixing exists.
Figures 2d) and 2e) show the resistances measured over the junction at fixed magnetic fields of $B=\pm$4~T. For these values the shift between the CNPs corresponds to a difference in filling factors of $\Delta\nu$=4. This leads to an overlap of filling factors in the two regions at certain backgate voltages, providing combinations of filling factors 2 and 6 in the unipolar cases as well as 2 and 2 in the bipolar case. As expected from equation (1), for the unipolar cases ($V_{BG}\approx$ -2 and 8~V) at positive magnetic field, $R_2$ shows a value of $1/3\cdot R_K$, while $R_1$ goes down to zero. In between, at $V_{BG}\approx$ 3~V for the bipolar case, equation 2 predicts $R_1=1 \cdot $R$_K$.
Figure 2e) shows the measurements at opposite magnetic field, resulting in a switching of the behavior of the two resistances. This indicates a change in the direction of the edge channels which can only be observed in four terminal experiments.
\begin{figure}[t]
\includegraphics[width=.95\columnwidth]{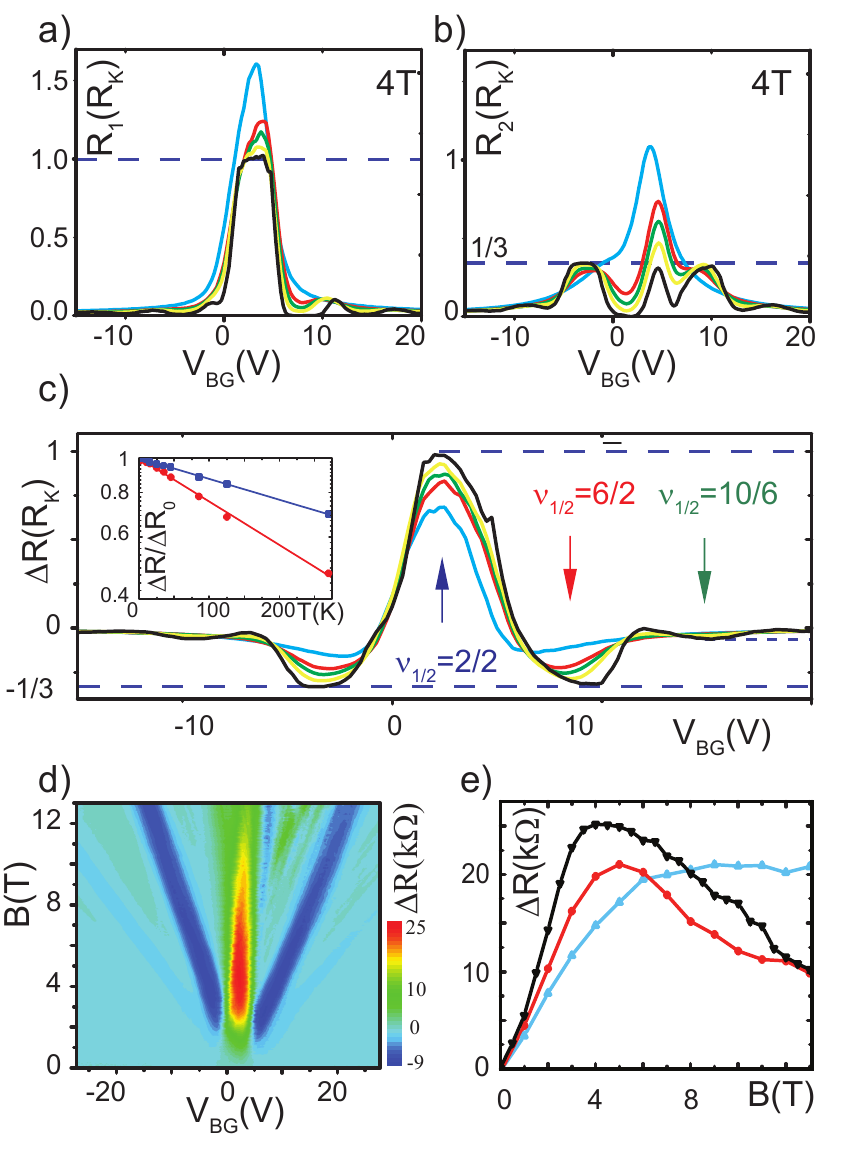}
\caption{\label{fig:bild3}a),b) The longitudinal resistances across the junction at $B$=4~T and different temperatures (1.5~K, 45~K, 85~K, 125~K, 270~K from lowest to highest peak resistance). The dotted blue lines indicate 1 and 1/3 of the von Klitzing constant. c) Difference $\Delta R = R_1-R_2$ for different temperatures as in a) and b). The arrows mark the voltages at which the normalized difference is obtained as plotted on a logarithmic scale as function of temperature in the inset. d) The difference $\Delta R$ as a function of backgate and magnetic field at $T$=1.5~K. e) The peak resistance in the p-n regime at $V_{BG}=2.5$~V is plotted for $T$=1.5~K (black), 125~K (red), and 270~K (cyan), as a function of magnetic field.}
\end{figure}\\
To study the interaction of the edge channels at the junction in detail, especially for the bipolar case, measurements are performed as shown in Fig. 3 with temperatures ranging from 1.5 to 270~Kelvin. While the resistance $R_1$, measured before the interaction of the counterpropagating edge states, shows some changes, a greater impact of the temperature can be seen at R$_2$, measured after the equilibration. For increasing temperature, the values after and before the interaction become more similar. To remove contributions of the sample geometry, the difference between the two measurements $\Delta R=R_1-R_2$ is analyzed (Fig. 3c). At low temperatures the quantizations at R$_K$, -1/3~R$_K$ and also -1/15~R$_K$, which corresponds to the combination of filling factors 6 and 10, can be identified. Although the peak heights deviate from the expected values at increasing $T$, the effect is still strong at high temperatures. This can be attributed to the stability of the quantum Hall effect in graphene even at high temperatures \cite{rtqh}, but also contains a contribution of the classical Hall effect due to different charge carrier densities in the two areas. The inset shows the normalized resistance values on a logarithmic scale as a function of temperature for the bipolar and the unipolar case as marked by the arrows in the main figure. Both values exhibit a monotonic decrease, highlighted with linear fits. Assuming an exponential behavior $\Delta R\propto exp(-k_BT/E_i)$, the according fits yield E=~64~meV for the bipolar case and 28~meV (31~meV) for the unipolar case with electrons (hole) as majority charge carriers. The two slopes show significantly different values for the two cases which can be attributed to the different energetic distances between Landau Levels (LL). In single layer graphene the energy of the LLs are given by $E_N=v_F\sqrt{2e\hbar BN}$, resulting in a lower distance between LL 1 and 2 with respect to 0 and 1. Since for the bipolar case only filling factor 2 is important, the distance is higher and therefore the influence of temperature, i.e. of LL broadening, is less compared to the unipolar case with filling factors 2 and 6. \\
Figure 3d) shows the difference $\Delta R$ in the resistances for a broad range of backgate voltages and magnetic fields. The quantizations for $\nu_{1/2}=$2/6, $\nu_{2/2}=$2/2, and $\nu_{1/2}=$6/2 are clearly visible. In contrast to the unipolar cases $\Delta R$ surprisingly decreases after $B$=4~T with increasing magnetic field for the bipolar case. The according values at a fixed backgate voltage of $V_{BG}$=2.5~V are shown in Fig. 3e) for three different temperatures. For $T$=1.5~K  $\Delta R$ exhibits a linear increase due to the different Hall voltages in the two areas. Around 4~T, the values quantizes at $R_K$ as explained before, indicating a full mixing of the edge states. For higher magnetic fields, it drops, suggesting a suppression of edge-channel equilibration at the p-n interface. Apparently, magnetic field localizes the edge channels strongly and suppresses the mixing of the counterpropagating edge states.
At higher temperatures, e.g. at $T=$270~K, the linear increase for low fields is flatter and also the quantization at $B=$4~T is not developed. For higher magnetic fields, $\Delta R$ does not drop but also does not follow the classical linear dependence. Interestingly, at the highest magnetic field $\Delta R$ grows by a factor of two with respect to the low temperature measurement. We attribute this observation to an increased mixing of the counterpropagating edge states indicating the transition to the classical behavior.\\\

In summary, a method is introduced which makes it possible to manipulate the doping level of single layer graphene in a defined region by AFM nanomachining. The demonstrated technique could be used to create graphene devices with small-sized and locally defined potential variations. Tuning a global gate, a junction of different charge carrier densities and polarities is created and studied using magnetotransport measurements. Quantized resistance values as expected from the edge-channel picture are observed as well as the switching of longitudinal resistances by the polarity of the magnetic field. The transport across the p-n junction shows an astonishing dependence of edge-channel equilibration on the magnetic field and temperature.\\\\ 
The authors acknowledge financial support by DFG Priority Program Graphene and the excellence cluster QUEST within the German Excellence Initiative.

\newpage
\end{document}